\documentclass[prl, twocolumn]{revtex4-1}
\usepackage[draft]{hyperref}
\usepackage{amsmath}
\usepackage{physics}
\usepackage{graphicx}	
\usepackage{siunitx}
\usepackage{ulem}

\begin{document}

\title{Coherent single-atom superradiance}

\author{Junki Kim}
\author{Daeho Yang}
\author{Seung-hoon Oh}
\author{Kyungwon An}
\email[Correspondence to: ]{kwan@phys.snu.ac.kr} 
\affiliation{School of Physics and Astronomy, Seoul National University, Seoul 08826, Korea}
	
\begin{abstract}
Superradiance is a quantum phenomenon emerging in macroscopic systems whereby correlated single atoms cooperatively emit photons.
Demonstration of controlled collective atom-field interactions has resulted from the ability to directly imprint correlations with an atomic ensemble.
Here, we report cavity-mediated coherent single-atom superradiance: single atoms with predefined correlation traverse a high-Q cavity one by one, emitting photons cooperatively with the $N$ atoms already gone through the cavity. 
Enhanced collective photoemission of $N$-squared dependence was observed even when the intracavity atom number was less than unity.
The correlation among single atoms was achieved by nanometer-precision position control and phase-aligned state manipulation of atoms by using a nanohole-array aperture. 
Our results demonstrate a platform for phase-controlled atom-field interactions.\\ \\
{\bf One-sentence summary:} 
Enhanced collective photoemission is observed from correlated atoms traversing an optical cavity.
\end{abstract}
\maketitle

Superradiance 
is a collective radiation phenomenon by a number of quantum emitters\cite{DIcke1954}. 
In the original prediction, exchange symmetry is present in closely packed emitters whose inter-particle distance is much smaller than the transition wavelength, and therefore dipole-dipole correlation emerges during their spontaneous decay process.
The correlation makes the ensemble behave collectively and induces enhanced interaction with the vacuum fields, 
leading to stronger and faster radiation emission compared to the ordinary spontaneous emission.
Early experiments performed
with a large number of emitters (as in a dense atomic vapor or in a beam)  
reported observations consistent with the prediction\cite{Gross1982,Skribanowitz1973}.
Recent technical advances 
have 
enabled the realization of superradiance in various systems such as a Bose-Einstein condensate\cite{Inouye1999}, quantum dots\cite{Scheibner2007} and trapped atoms coupled to a cavity\cite{Reimann2015}.


The mutual phase correlation among atoms is the key to superradiance. It can make the ensemble behave as a single macro dipole. 
Moreover, direct control of atomic phases enables controllable collective atom-field interactions.
In recent experiments, the phase of atoms in an ensemble was imprinted by a single photon pulse\cite{Scully2009a,Rohlsberger2010,Roof2016} or a frequency-swept laser pulse\cite{Norcia2016}.
The ensemble then started superradiant emission without a threshold or an initial time delay.
The output field in this case follows the given imprinted phase and thus its spatial mode overlaps with the input mode, making it hard to distinguish the input and output fields spatially. 
This approach works only in the pulsed regime.
Observation of a superradiant state in a Bose-Einstein condensate couple to a cavity was another notable work\cite{Baumann2010}. However, it relied on self-organization of atoms based on a thermodynamic principle, and thus further tunability could not be attained.

Another approach to achieve controllable superradiance is to prepare emitters in a cavity and to manipulate the quantum state of individual emitters. Ions\cite{Casabone2015}, neutral atoms\cite{Neuzner2016} and artificial atoms based on superconducting circuits\cite{Mlynek2014} 
have been used in this approach.
The results include 
immediate strong and fast radiation emission 
and controllability between superradiance and subradiance. 
However, technical difficulties have limited the number of emitters involved in the superradiance only up to two.

We present 
an approach to realize phase-controlled superradiance 
whereby single atoms are prepared in the same quantum superposition of ground and excited states traverse a cavity one by one. 
The long-lived cavity field then mediates collective interaction among the phase-correlated single atoms separated in time, leading to superradiance.
The collective interaction is one-sided 
in that the emission of a particular atom in the cavity is cooperative only with the preceding atoms.
Even when at most only one atom is present in the cavity, tens of atoms participate in the superradiance and the emission intensity is proportional to the square of the number of the participating atoms.




\begin{figure*}

\centering\includegraphics[width= 0.85\textwidth]{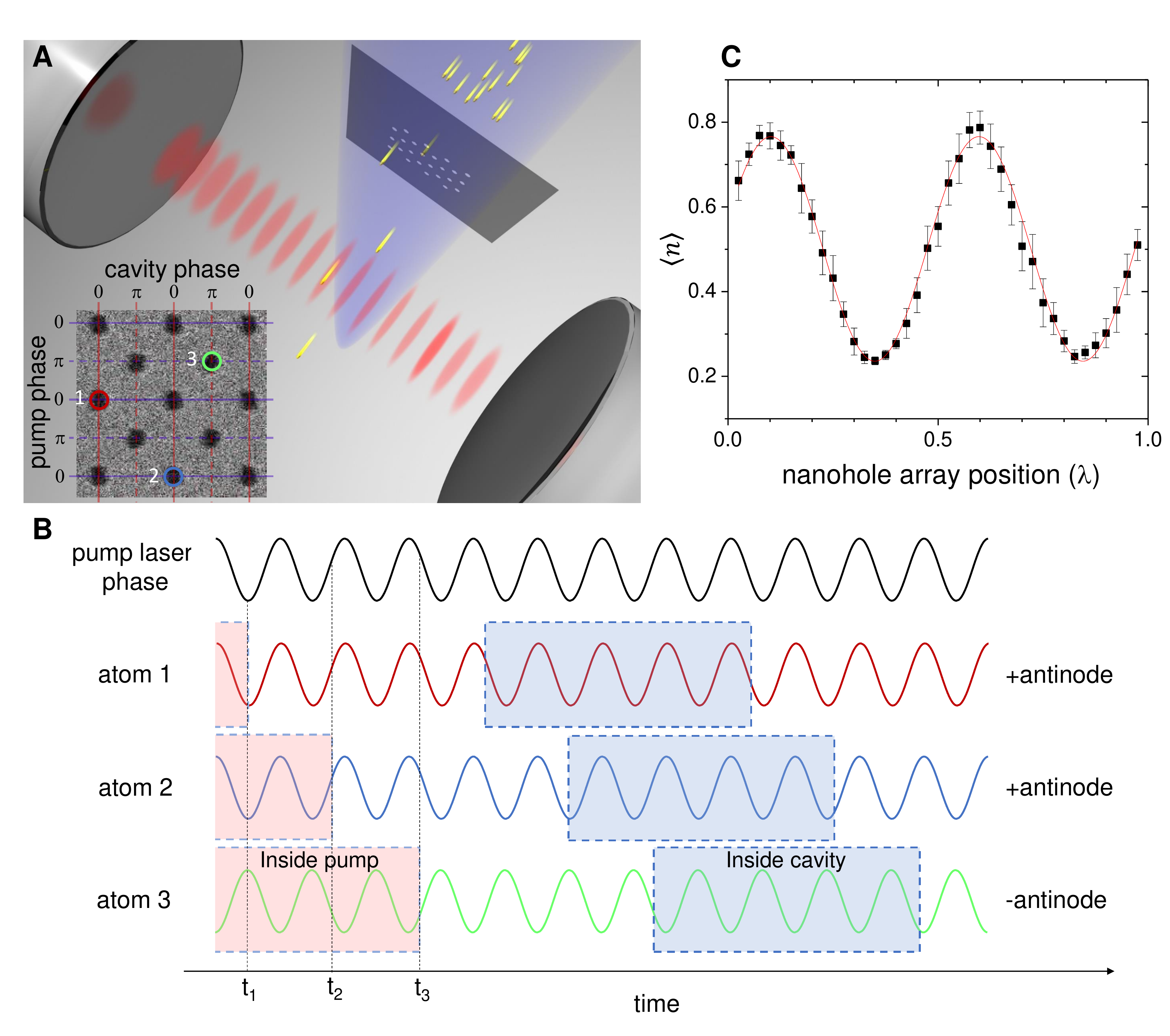}
\caption{ 
	\textbf{Phase-controlled atom-cavity interaction with a nanohole-array aperture.}
	({\bf A}) Barium atoms (yellow flying spheres) go through a nanohole-array aperture and are prepared in a desired state by a pump laser (blue beam).  The atoms interact with the cavity field in a given interaction time $\tau$ and emit photons into the cavity mode in both collective and non-collective ways.
	The inset shows a focused-ion-beam image of the nanohole-array aperture \cite{mm}.
	The nanohole array dimension is 25$\lambda$ by 25$\lambda$, spanning 19.8$\mu$m both horizontally and vertically. Here $\lambda$=791nm is the atomic transition wavelength.
	({\bf B}) The atom-cavity relative phase is prescribed by the array structure. Consider three atoms going through nanoholes 1, 2 and 3 as indicated in {\bf A} and then exiting the pump field at time $t_1, t_2$ and $t_3$, respectively. The pump laser phase at one of the zero-phase planes is shown with the imprinted atomic phases.
	Although atom 3 has the opposite imprinted phase to the others, 
	it goes through an antinode opposite in phase to the antinodes that atoms 1 and 2 go through.
	As a result, all three atoms would have identical atom-cavity relative phases.
	({\bf C}) Varying the nanohole-array position along the cavity axis changes the atomic trajectory from crossing nodes to crossing antinodes, and therefore the cavity mean photon number (black square) varies sinusoidally when pumped by fully excited atoms. The signal visibility is 0.54 and the corresponding effective atomic distribution has a full width at half maximum of 0.29$\lambda$. Red solid line is a sinusoidal fit to the data. 
	}
\label{fig:scheme}
\end{figure*}

Our system, adapted from Ref. \cite{Lee2014}, consists of a supersonic barium atomic beam and a high-Q optical cavity which the atoms resonantly interact with (Fig.~1A). 
The barium-138 atoms are prepared in a superposition state of the ground and excited states just before they enter the cavity mode by a pump laser propagating perpendicular to the cavity axis as well as to the atomic beam direction.
The atomic phase imprinted by the pump laser depends on the position at which the atom traverses the pump laser.
The phase of the atom-cavity coupling also alternates 0 and $\pi$ radian following the standing wave structure of the cavity mode.
A checkerboard-pattern nanohole array 
is used
as an atomic beam aperture in order to localize and control the atomic position.
The localized atoms then selectively pick up the phase of the pump laser as well as the cavity field corresponding to their positions prescribed by the array structure. 
As a result, the atom-field relative phase is the same for every atom traversing the cavity 
(Fig.~1B). 
The desired atomic internal state is prepared by the pump laser with a pulse area of $\Theta = \int[\Omega_p (x)/v] dx$, where $\Omega_p(x)$ is the Rabi frequency due to the pump laser and $v$ is velocity of the atom. The atomic state can then be expressed as $\left|\psi_{\rm{atom}}\right\rangle = \sin(\Theta/2)\left|\rm{e}\right\rangle + \cos(\Theta/2)e^{i\phi}\left|\rm{g}\right\rangle$, where $\phi$ is the atomic phase imprinted by the pump laser.
Atomic correlation between any two of the injected atoms is then given by $\langle \sigma_i^\dagger \sigma_j\rangle= \frac{1}{4} \sin^2 \Theta$,
where $\sigma_i=\left|{\rm g}\right\rangle \left\langle {\rm e}\right|$ is the lowering operator of the $i$-th atom, showing 
that the atom-atom correlation is maximized when $\Theta = \pi/2$.

Injected atoms then emit photons into the cavity mode and build up the cavity field. 
A previous study assuming a lossless cavity expected enhanced collective emission by consecutively injected $N$ atomic dipoles to show explicit $N^2$ dependence\cite{Kien1993}. 
The longlasting cavity field links the atoms together and the expected photon number is exactly the same as that of simultaneously injected $N$ dipoles (see Fig.\ S1).

When a cavity has a finite decay, the gain (emission by atoms) and the loss (absorption by atoms as well as the cavity decay) of the cavity field would be balanced in its steady state. The averaged cavity photon number $\langle n \rangle$ in the steady state can be obtained from the quantum master equation (see Supplementary Text Section 2.2 for details)
and it is approximately given by
\begin{equation}
    \langle n\rangle \approx \frac{ \langle N_c\rangle \rho_{\rm{ee}} (g\tau)^2}{2-(2\rho_{\rm{ee}}-1)\langle N_c\rangle (g\tau)^2} + (\langle N_c\rangle |\rho_{\rm{eg}}|  g\tau)^2.
\label{n_expression}
\end{equation}
\noindent where $\langle N_c\rangle \equiv r/\gamma_c$ is the mean number of atoms injected into the cavity during the cavity-field
decay time $1/\gamma_c$ with $r$ the atomic injection rate, $\rho_{\rm{ee}}$ and $\rho_{\rm{eg}}$ are the density matrix elements of atomic state with the subscripts `e' and `g' represent excited and ground states, respectively, $g$ is the atom-cavity coupling constant and $\tau$ is the atom-cavity interaction time.

The first term, approximately proportional to $\frac{1}{2} \langle N_c\rangle$ when $\langle N_c\rangle (g\tau)^2\ll 1$, is due to the non-collective emission of atoms, including spontaneous and stimulated emission as well as the cavity-QED effect. 
The second term, exhibiting a quadratic dependence on $\langle N_c \rangle$, is due to collective emission, \textit{i.e.} the superradiance.
Compared to the case with a lossless cavity\cite{Kien1993}, the number of atoms participating in the superradiance is identified to be $\langle N_c\rangle$ in our case (see Supplementary Text Section 2.1). 
When $\langle N_c\rangle \gg 1$, the second term dominates the emission and the field state  approximately becomes a coherent state $\left|\alpha\right\rangle$ with $\alpha = -i \langle N_c\rangle \rho_{\rm eg} g\tau$. 

The mean intracavity atom number $\langle N\rangle$ is related to $\langle N_c\rangle$ as $\langle N_c\rangle = \langle N \rangle /(\gamma_c\tau)$.
If the cavity-field decay time $1/\gamma_c$ is much larger than $\tau$ ($\gamma_c\tau \ll1$), $\langle N_c\rangle$ can be much greater than unity even when the mean intracavity atom number $\langle N\rangle$ is less than unity and thus the collective effect can take place. 
The cavity field mediates the collective behavior among the time-separated $\langle N_c\rangle$ atoms that are going through the cavity individually during the cavity-field decay time, 
leading to the single-atom superradiance. 
Around 22 atoms are involved in the collective emission when a single atom is present in the cavity mode on average.


In our experiment, the phase-aligned atomic dipoles prepared with the aforementioned nanohole-array were injected into the cavity 
and the mean intracavity photon number in the steady state was measured with a single-photon-counting module.
The atom-cavity interaction was in the strong coupling regime with (g,$\gamma$, $\gamma_c$) = 2$\pi \times$ (290, 25, 75)kHz, where $g$ is the atom-cavity coupling averaged over the atomic distribution centered around the antinodes of the cavity and $\gamma$ ($\gamma_c$) is the atomic polarization (cavity-field) decay rate. 
The single-atom cooperativity was $C = g^2/\gamma\gamma_c = 44$.
The mean travel time of atoms from the pump to the cavity field was about 200ns whereas the mean atom-cavity interaction time $\tau = 101$ns.
As a comparative counterpart, we also performed the experiment with a 250$\mu$m$\times$25$\mu$m-sized rectangular atomic beam aperture for the case of atoms with random phases. In the latter case, the atomic beam was injected into the cavity mode with a small tilt angle in order to induce Doppler shifts so as to achieve a uniform atom-field coupling\cite{Choi2006,Yu2006,An1997b,Hong2012}, whose strength is a half of the maximum coupling strength.

\begin{figure*}
	\center \includegraphics[width=1\textwidth]{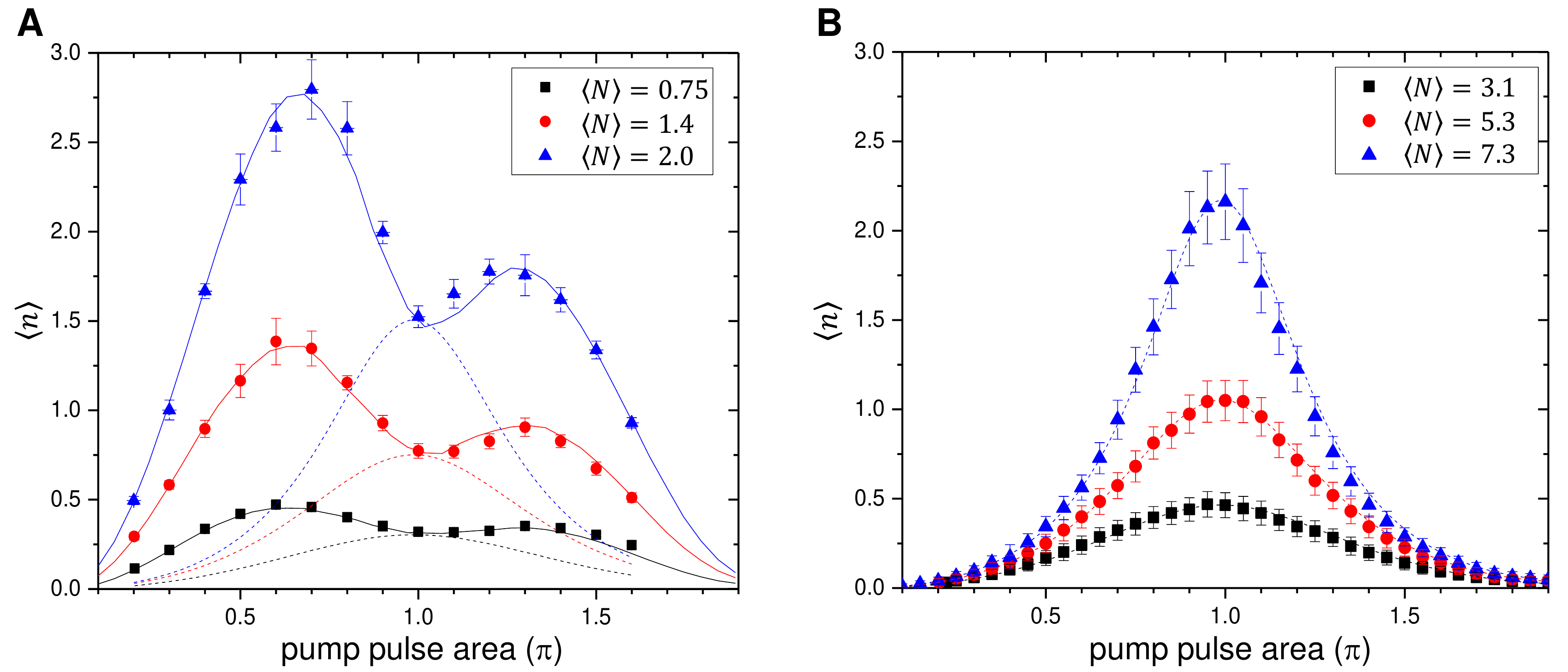}
	\caption{
		\textbf{Intracavity photon number dependence on the atomic state.}
		The intracavity mean photon number $\langle n\rangle$ versus the pump pulse area $\Theta$ for various mean atom numbers $\langle N\rangle$ in the cavity with ({\bf A}) phase-aligned atoms and ({\bf B}) atoms with random phases. Vacuum Rabi angle $g\tau$ is associated with the cavity-enhanced spontaneous emission probability $\sin^2(g\tau)$ during the atom-cavity interaction time $\tau$ for a single atom. The value of $g\tau$ is 0.18 radian for the phase-aligned atoms, averaged over the finite atomic distribution around the antinode of the cavity field, and 0.10 radian for the random phase atoms with a traveling-wave atom-cavity coupling. Solid (dashed) lines are theoretical predictions for the case of phase-aligned (random phase) atoms with the actual experimental parameters \cite{mm}.}
	\label{fig:pump}
\end{figure*}

\begin{figure*}
	\center \includegraphics[width=1\textwidth]{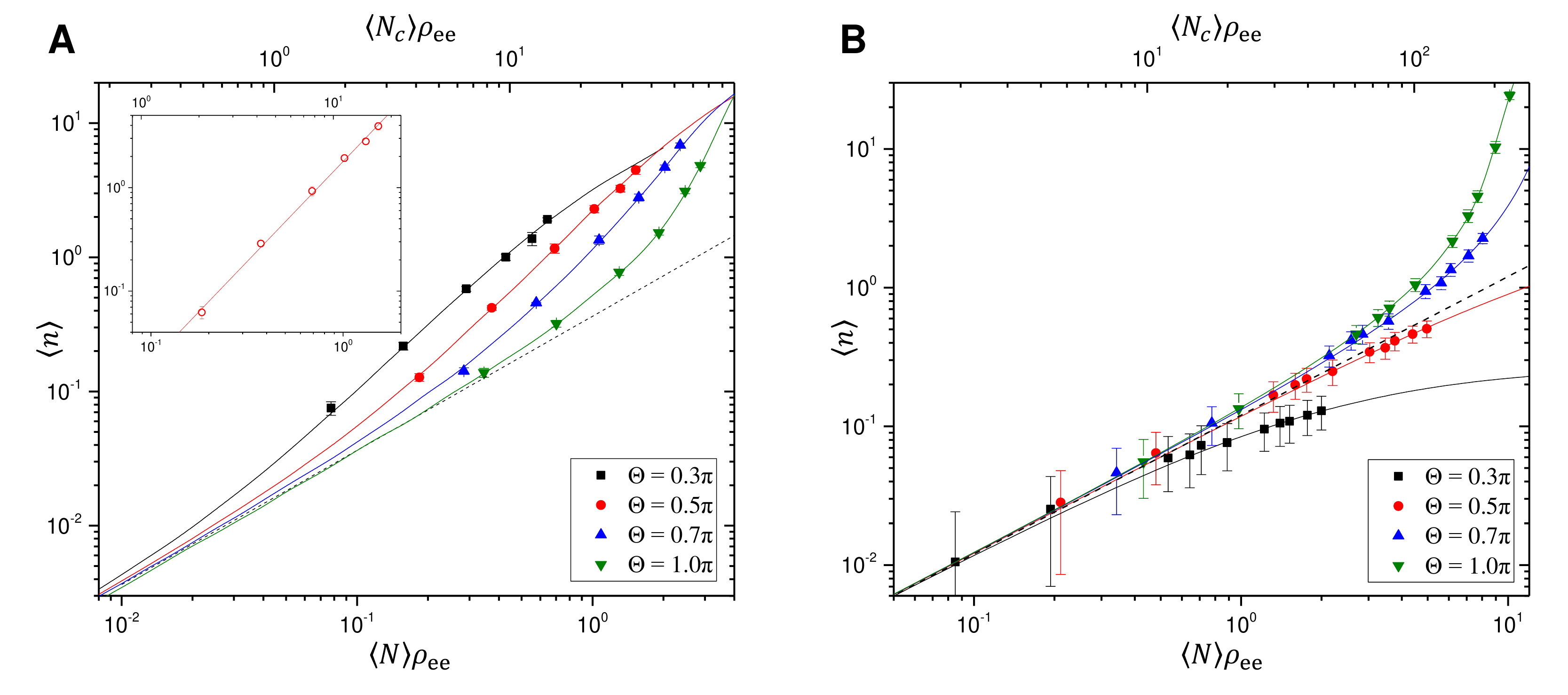}
	\caption{
		\textbf{Intracavity photon number dependence on the number of atoms.}
		The intracavity mean photon number $\langle n \rangle$ versus the excited-state mean atom number $\langle N \rangle\rho_{\rm{ee}}$ for ({\bf A}) the phase-aligned case and ({\bf B}) the random phase case. We use $\langle N \rangle\rho_{\rm{ee}}$ instead of just $\langle N \rangle$ for the horizontal axis in order to align the non-collective emission contribution as a common base line. The inset shows $\langle n \rangle$ with the non-collective contribution subtracted for $\Theta=0.5\pi$ cases and a linear fit to data with the log-log slope of $1.94\pm0.04$. Dashed lines correspond to the expected photon number made only by the cavity-enhanced spontaneous emission of atoms. Solid lines are theoretical predictions with the actual experimental parameters \cite{mm}.
		}
	\label{fig:Nvn}
\end{figure*}

The collective emission described by the second term in Eq.~(1) 
is expected to have the quadratic dependence on two parameters, the induced atomic dipole moment $\propto|\rho_{\rm{eg}}|$ and the atom number $\langle N_c \rangle$. 
First, we investigated  
$|\rho_{\rm{eg}}|$ dependence of collective emission by varying the pump pulse area (Fig.~2). 
Due to the relation $|\rho_{\rm{eg}}| = |\frac{1}{2}\sin\Theta |$ for the prepared superposition state, the atomic dipole moment would be maximized with equal ground- and excited-state populations ($\Theta = 0.5\pi \,\;\rm{or}\, 1.5\pi$), and so would be the collective emission.
Clear enhancement was observed when the atoms are prepared in the phase-aligned superposition states. The enhancement was more than ten-fold for $\Theta<0.3\pi$ (also see Fig.\ S2).
Combined contributions by $\rho_{\rm ee}$ (non-collective) and $|\rho_{\rm eg}|$ (collective) make $\langle n \rangle$ maximized near $\Theta \simeq 0.7\pi$.
Due to the small overlap between the pump laser field and the cavity mode (both are Gaussian), 
the collective emission process is somewhat disturbed by the stray pump field in the cavity
when the pump intensity is strong, resulting in the enhancement reduction for $\Theta > \pi$.
On the other hand, in the case of random phase, the photon number is given by the non-collective emission only, and thus it is maximized with fully inverted atomic states ($\Theta = \pi$).

The enhancement is strongly dependent on the atomic phase purity. In reality, there are several sources of phase noise. 
Finite atomic localization sets the lower bound of atomic phase variance. 
Atomic spontaneous emission into free space also contributes to phase diffusion of atoms, reducing $|\rho_{\rm{eg}}|$ by 6\%.
In addition, the pump laser has a phase uncertainty: the laser phase diffuses in time with a finite laser linewidth.
If we intentionally make the pump laser linewidth larger, the superradiant enhancement becomes smaller (see Fig.\ S3). 
We performed quantum-trajectory simulation as well as quantum master equation calculation with the experimental parameters and our data well agree with the numerical results \cite{mm} (also see Fig.\ S4).

Figure 3 
shows the mean intracavity photon number $\langle n \rangle$ versus the excited state atom number $\langle N \rangle\rho_{\rm ee}$.  
When the atoms have no dipole moment (Fig.\ 3B), 
only the non-collective emission is present. With a small number of atoms, the cavity field is mainly made by spontaneous emission of atoms (dashed line) and its photon number increases linearly to the atom number. As the accumulated photon number gets larger, stimulated emission and absorption become dominant over the spontaneous emission and the system lases for positive inversion ($\rho_{\rm{ee}} - \rho_{\rm{gg}} > 0$) or the photon number plateaus for negative inversion ($\rho_{\rm{ee}} - \rho_{\rm{gg}} <0$). Especially for positive inversion, a rapid growth of the photon number starts to occur at $\langle n \rangle \simeq 1$, which is the well-known lasing threshold in the conventional lasers\cite{Bjork1994}. 

However, when atoms have the same phase (Fig.\ 3A), 
photon emission is enhanced nonlinearly with its log-log slope getting steeper than unity. The measured intracavity photon numbers are consistently larger than the photon number made only by the cavity-enhanced spontaneous emission (dashed line). When the pump pulse area is $0.5\pi$, corresponding to $\left| \psi_{\rm{atom}} \right\rangle \simeq (\left|\rm{e}\right\rangle + e^{i\phi}\left|\rm{g}\right\rangle)/\sqrt{2}$, the observed log-log slope is 1.66$\pm$0.01. 
After subtracting the contribution by the non-collective emission corresponding to the dashed line,
the recalculated log-log slope becomes 1.94$\pm$0.04 (see the inset of Fig.\ 3), which indicates the observed emission is dominantly superradiance proportional to the square of the number of atoms. 
A near-quadratic growth appears even in the negative inversion case of $\Theta = 0.3$, in which only 21\% of atoms are in the excited state with the rest in the ground state.
When $\Theta >0.5\pi$, the atoms have positive population inversion and thus the photon number grows further by stimulated emission beyond the level by the collective emission. In this case, it is impossible to isolate the collective emission effect clearly in the log-log plot.

It is also notable that the log-log slope is almost invariant for a large range of $\langle N_c \rangle$ for $\Theta \leq 0.5\pi$. The theory expects that
the quadratic dependence on $\langle N_c \rangle$ 
would be dominant 
in the region of $(1+\cos\Theta)^{-1} < \langle N_c \rangle<(g\tau)^{-2}$ for the perfectly phase-aligned atoms although the practical phase noise would make the domain somewhat reduced. 
Such a broad-range quadratic growth, occurring independently of $\langle n \rangle$ values, including $\langle n \rangle \ll 1$ as in Ref.\cite{Bohnet2012},
is a distinctive feature of the present superradiance compared to the drastic slope change occurring near the threshold condition of $\langle n \rangle \simeq 1$ in the ordinary lasing case.
The absence of the usual lasing threshold or thresholdless lasing in the present superradiance cannot be explained in terms of the  so-called $\beta$-factor in ordinary lasers based on non-collective emission\cite{Khajavikhan2012}. In our case $\beta=(g\tau)^2\simeq 0.034$ in the nanohole-array-aperture case (Fig.~2A and 3A) and 0.011 in the rectangular-aperture case (Fig.~2B and 3B) (see Supplementary Text Section 2.3). The latter is consistent with the large mean photon number change occurring at the threshold in Fig.~3B ($\Theta>\pi/2$).
Note also that the range of superradiance or
the maximum number of atoms participating in the collective emission can be easily scaled up by choosing smaller $g\tau$ values (see Fig.\ S5).
This feature may provide a new approach in building thresholdless lasers.

The present single-atom superradiance can be viewed as a consequence of one-sided 
interaction among a series of atoms separated by tens of meters. Note that the photon emitted by a preceding atom interacts with the next atom after traveling $c\tau/\langle N \rangle$ (about 30m for $\langle N \rangle = 1$) when we unfold mirror refections although their average distance in real space is only hundreds of micrometers.
Due to causality, only the preceding atoms can then affect the quantum states of the following atoms. 
This 
interaction 
induces the emission rate of the atom in the cavity to be twice larger than the emission rate per atom in the usual superradiance  
(see Supplementary Text as well as Fig.\ S6).
The time-separated atoms linked by such one-sided interaction can form 
atom-atom interaction systems, which can serve as a testbed for various quantum many-body physics\cite{Lodahl2017}.

The present study deepens our understanding on matter-light collective interaction and provides a new insight on the field-mediated long-range\cite{Douglas2015,Meir2014} 
interactions. 
In addition, the phase-controlled many-atom-field interaction based on the nanohole-array technique can be used in non-classical field generation such as optical Schr\"{o}dinger cat states and highly-squeezed vacuum states\cite{Yang2016}, even in a lossy cavity contrary to the previous studies in the microwave region\cite{Deleglise2008}, 
as well as in realizing superabsorption\cite{Higgins2014}.
The greatly enhanced single-atom emission may be useful in constructing efficient quantum interfaces\cite{Hammerer2010}. 

\hbox{}

\noindent\textbf{Acknowledgements}

\noindent
We thank Moonjoo Lee and Hyun-Gue Hong for helpful comments. This work was supported by a grant from Samsung Science and Technology Foundation under Project No. SSTF-BA1502-05.
The authors declare no competing financial interests.
All data needed to evaluate the conclusions in the paper are present in the paper and/or the Supplementary Materials.\\

\noindent\textbf{List of Supplementary Materials}\\
Materials and Methods\\
Supplementary Text\\
Figs. S1 to S6\\
References (\textit{32-35})
%
%
%
%

%
%

\setcounter{equation}{0}
\widetext
\newpage
\newpage

\begin{center}
{\huge \bf Supplementary Materials}
\end{center}
\hbox{}
\hbox{}

\noindent{\bf \Large 1 Materials and methods }

\hbox{}

\noindent{\bf \large 1.1 A nanohole-array aperture, a supersonic atomic beam and a high-Q cavity}

\hbox{}

Our nanohole-array is fabricated on a silicon nitride membrane (Si$_3$Ni$_4$) of 10nm thickness by using the focused-ion-beam technique. The distance between holes in the pump beam direction as well as in the cavity axis directions is equal to the atomic transition wavelength $\lambda=791$nm.
The nanohole diameter is 0.24$\lambda$. 
Actual atomic localization width would be slightly larger than the nanohole diameter because of a finite atomic beam divergence and a small angular mismatch between the array and the cavity axis. 
The atomic distribution in the cavity is estimated to have a full-width at half-maximum (FWHM) of 0.29$\lambda$ for each nanohole, 
which is obtained from the observed contrast of the modulation of the cavity photon number as the array position along the cavity axis 
while the excited atoms are injected into the cavity through the nanohole-array.
The nanohole array dimension is 25$\lambda$ by  25$\lambda$, spanning 19.8$\mu$m both horizontally and vertically. The vertical dimension is much smaller than the cavity-mode full width (83.8$\mu$m), ensuring near constant atom-cavity coupling constant in that direction.

To investigate the single-atom superradiance effectively, we need to satisfy the following conditions: negligible frees space decay of atoms during the atom-field interaction ($\gamma\tau\ll$1), a small single-atom vacuum-Rabi angle to suppress the gain saturation effect ($g\tau\ll1$) (\textit{18,19}) 
and a sufficiently long cavity-field decay time ($\gamma_c\tau\ll1$). 
We use barium-138 having a narrow spin-forbidden transition ($^1$S$_0$ -- $^3$P$_1$) of $\gamma = 2\pi \times 25$kHz natural linewidth (half width) and zero nuclear spin.
We treat the atoms as a two-level system with $^1$S$_0$ being the ground state and $ ^3$P$_1$ the excited state.
We use a supersonic beam of barium atoms with a fast mean velocity of $v_{\rm{mean}} = 755$ m/s and a narrow velocity distribution with a FWHM of $0.24v_{\rm{mean}}$. 
A bias magnetic field is applied along the atomic beam direction so as to define the quantization axis of the atoms and to lift the magnetic sublevel degeneracy.
The polarizations of the pump field and the cavity probe field are also set to parallel to the atomic beam direction in order to access the magnetic-field-insensitive $\pi$ transition ($\Delta m = 0$) of atoms.
The cavity is made of supermirrors with a finesse of $0.92\times10^6$ and a radius of curvature of 10.0cm. The cavity length is 1.09mm and the mode waist is 42.5$\mu$m in the direction of the atomic beam and 41.9$\mu$m in the direction of the pump beam.

\hbox{}

\noindent{\bf \large 1.2 Success rate of atomic state manipulation}

\hbox{}

Atomic state is prepared by a pump laser. However, the velocity distribution of the atomic beam and the free space spontaneous emission decay of the excited state induce small imperfection in state preparation. To verify resulting atomic state after passing through the pump laser, we measured the ground state population from the fluorescence
of the 553nm cycling transition ($^1$S$_0\leftrightarrow ^1$P$_1$). Under the $\pi$-pulse condition ($\Theta = \pi$) of the pump, the observed population in the excited $^3$P$_1$ state was 93$\pm$1\%, well-agreeing with the numerical simulation result of 93.6\%.

\hbox{}

\noindent{\bf \large 1.3 Photon number calibration and atom number measurement}

\hbox{}

The mean photon number $\left<n\right>$ in the cavity is calibrated by using the photon number clamping effect. When the cavity is pumped by the atoms with a negative inversion and random phases, $\left<n\right>$ would converge to $\rho_{\rm{ee}}/(1-2\rho_{\rm{ee}})$ as $\left<N_c\right>\rightarrow \infty$ (detailed in Section 2.2.1), 
only depending on the initial excited state population, which can be independently obtained with the aforementioned fluorescence measurement technique. We found that a single intracavity photon in the steady state corresponds to $1.1\times10^5$ counts per second measured by our single photon counting module. 

Once the photon number is calibrated, we can extract the mean intracavity atom number $\left<N\right>$ by comparing the measured $\left<n\right>$ in the case of fully excited atoms ($\Theta = \pi$) with the result of the master equation calculation. The calibration result was verified to be reasonable by performing a linear regression analysis between the estimated $\left<N\right>$ and the directly measured atomic fluorescence (at 553nm) in the large atom number region (with the adjusted coefficient of determination of $R^2 = 0.94$).

\hbox{}

\noindent{\bf \large 1.4 Numerical simulations of experiment}

\hbox{}

To compare the experimental results with theoretical expectations quantitatively, we performed two different numerical studies: the master equation calculation and the quantum trajectory simulation (QTS) (\textit{32,33}). 
QTS appropriately treats transient fluctuations including phase diffusion and atom number fluctuation but requires high computational power. Therefore, we performed QTS for the phase-aligned atom case, which has less intracavity atoms, and the master equation approach for the random phase case. The master equation for our experiment is fully derived in Section 2.2. The master equation was solved for the steady state. With the same simulation parameters, two methods perfectly agree with each other.

\hbox{}
\hbox{}

\noindent{\bf \Large 2  Supplementary text}

\hbox{}

\noindent{\bf \large 2.1 Qualitative description of the single-atom superradiance}

\hbox{}


A two-level atom can be represented as a spin-$1/2$ particle and any quantum state of their ensemble can be expressed in terms of $\left|J,M \right\rangle$, eigenstates of total angular momentum operators $\sigma_\Sigma^2 = (\sum_i \vec{\sigma_i}^2)$ and $\sigma_{\Sigma,z}= \sum_i \sigma_{i,z}$.
The spontaneous radiation rate of the state is then given by 
\begin{equation}
	\Gamma_a \left<\sigma_\Sigma^\dagger\sigma_\Sigma\right> = \Gamma_a \sum_{i} \left<\sigma_i^\dagger \sigma_i\right> + \Gamma_a \sum_{i\neq j} \left<\sigma_i^\dagger \sigma_j\right>
\end{equation}
\noindent where $\Gamma_a$ is atomic natural linewidth,  $\sigma_\Sigma= \sum_i  \sigma_i$ and $\sigma_i$($\sigma_i^\dagger$) is the lowering (raising) operator for the $i$-th atomic state.

The first term of the right hand side represents the sum of excited state probabilities, corresponding to the total energy quanta carried by the atomic ensemble, and the second term is related with the mutual correlation among atoms. When the atoms have no correlation at all, atoms decay independently and no collective effect emerges.

In the original idea of superradiance, an atomic ensemble has a perfect exchange symmetry and it is thus expressed in terms of the eigenstates with the maximum total angular momentum $J=N/2$. Their spontaneous decay rates are given by

\begin{equation}
\Gamma_a\left<\sigma_\Sigma^\dagger\sigma_\Sigma\right>= \Gamma_a \left(\frac{N}{2} + M\right) \left(\frac{N}{2} - M + 1\right),
\end{equation}

\noindent exhibiting the most enhanced decay rate of $\Gamma_a \frac{N}{2} \left(\frac{N}{2} + 1\right)$ when $M=0$, corresponding to the brightest superradiant state.

On the other hands, when the atoms are in the same superposition state of $C_e |{\rm e}\rangle + C_g |{\rm g}\rangle$, the total ensemble can be expressed as $|\Psi_{\rm atom}\rangle = \prod_i(C_e |{\rm e}_i \rangle + C_g |{\rm g}_i \rangle)$. This state also satisfies the exchange symmetry and thus can be decomposed into the sum of the eigenstates,

\begin{equation}
	|\Psi_{\rm atom}\rangle = \sum_{k=0}^N C_{\rm e}^{N-k} C_{\rm g}^k \left(\begin{array}{cc}
N \\ k 
	\end{array}\right) \left|J=\frac{N}{2}, M=\frac{N}{2} - k\right\rangle.
\end{equation}
\noindent The spontaneous decay rate of this state is then given by

\begin{eqnarray}
	&&\Gamma_a\sum_k|C_{\rm e}|^{2(N-k)} |C_{\rm g}|^{2k} \left( \begin{array}{cc}	N \\ k	\end{array} \right) (N-k)(k+1)\nonumber\\
	&&\quad=\Gamma_a N(N-1)|C_{\rm e}|^2 |C_{\rm g}|^2 \sum_k |C_{\rm e}|^{2(N-k-1)} |C_{\rm g}|^{2(k-1)} \left( \begin{array}{cc}	N-2 \\ k-1	\end{array} \right)\nonumber\\
	&&\quad\quad+\Gamma_a N|C_{\rm e}|^2 \sum_k |C_{\rm e}|^{2(N-k-1)} |C_{\rm g}|^{2k} \left( \begin{array}{cc} 	N-1 \\ k 	\end{array} \right)\nonumber\\
	&&\quad=N(N-1) \Gamma_a |\rho_{\rm{eg}}|^2 + N \Gamma_a \rho_{\rm{ee}},
\end{eqnarray}
\noindent also exhibiting $N^2$-dependence of the superradiance. 
The photon emission rate per atom is $\rho_{\rm{ee}}\Gamma_a  + (N-1) |\rho_{\rm{eg}}|^2 \Gamma_a $, interpreted as the rate sum of  non-collective emission and collective emission with other $N-1$ atoms. When the ensemble is strongly coupled to a cavity with a coupling constant $g$ during a definite interaction time $\tau$, the emission rate becomes $ \rho_{\rm{ee}} g^2\tau + (N-1) |\rho_{\rm{eg}}|^2 g^2\tau$,  where the free-space decay rate $\Gamma_a$ is replaced with the cavity-assisted decay rate $g^2\tau$.

Such collective emission can also appear when phase-aligned atoms do not interact with a field simultaneously, as in our single-atom superradiance. From the numerical calculation results on the consecutive injection of numerous atoms into a lossless cavity field, we have inductively found that the emission rate of the \textit{N}-th atom is $ \rho_{\rm{ee}} g^2\tau + 2(N-1) |\rho_{\rm{eg}}|^2 g^2\tau$. Here the preceding $N-1$ atoms influence the emission of $N$-th atom and the photons already emitted by the former ones induce the stronger collective emission, leading to a twice larger collective emission rate than that in the ensemble case (see also Fig. S1).

When the cavity field dissipation is introduced, the field-mediated interaction with the preceding atoms is also reduced by $\exp(-\gamma_c \Delta t)$, where $\gamma_c$ is the cavity field decay rate and $\Delta t$ is the time elapsed since the preceding atom left the cavity. The effective number of atoms participating in the collective emission with the atom introduced in the cavity at time $t_0$ is then given by
\begin{equation}
	N_{\rm eff} = \int_{-\infty}^{t_0} P(t) \exp\left[-\gamma_c (t_0-t)\right] dt
\end{equation}
\noindent where $P(t)$ is the injection rate of atoms at time $t$. The resulting emission rate per atom is now $ \rho_{\rm{ee}} g^2\tau + 2 N_{\rm eff} |\rho_{\rm{eg}}|^2 g^2\tau$. 
When the atoms are randomly injected into the cavity field, $P(t)$ is equivalent to the mean atomic injection rate $r$ and the corresponding $N_{\rm eff}$ is $r/\gamma_c\equiv\left<N_c\right>$. For regular injection of atoms with a constant time interval between them, $P(t) = \sum_i \delta(t-i/r)$ and 
\begin{equation}
	N_{\rm eff} = \sum_{i=1}^\infty \exp(-i\gamma_c/r) = \frac{1}{\exp(1/\left<N_c\right>) - 1 },
\end{equation}
\noindent which can be also approximated by $\left<N_c\right>$ for $\left<N_c\right>\gg 1$.

In the steady state of the cavity field, the loss $2\gamma_c\left<n\right>$ by cavity dissipation and the gain $r P_e$ from atomic emission should be balanced. Here, $P_e$ is the photo-emission per atom in the cavity, given by the emission rate per atom times the interaction time. Therefore, the cavity photon number in the steady state is given by

\begin{eqnarray}
	\left<n\right> &=& \frac{r P_e}{2\gamma_c}= \frac{r}{2\gamma_c} \left[\rho_{\rm{ee}} (g\tau)^2 + 2 N_{\rm eff} |\rho_{\rm{eg}}|^2 (g\tau)^2 \right]\nonumber\\
	&=& \frac{1}{2} \left<N_c\right>\rho_{\rm{ee}} (g\tau)^2 + \left<N_c\right> N_{\rm eff} |\rho_{\rm{eg}}|^2 (g\tau)^2\nonumber\\
	&\approx& \frac{1}{2} \left<N_c\right>\rho_{\rm{ee}} (g\tau)^2 + \left<N_c\right>^2 |\rho_{\rm{eg}}|^2 (g\tau)^2.
\end{eqnarray}

\hbox{}

\noindent{\bf \large 2.2 Quantum master equation approach}

\hbox{}

The time evolution of the cavity field can be fully described by the quantum master equation. The equation is based on the atom-field interaction governed by the Jaynes-Cummings Hamiltonian, $\hat{H} = \hbar\omega_c\hat{a}^\dagger \hat{a} + \hbar\omega_a\hat{\sigma}_z+\hbar g (\hat{a}\hat{\sigma}^\dagger+\hat{a}^\dagger\hat{\sigma})$, and the stochastic decay described by the Lindblad equation. 
Here $\hat{a} (\hat{a}^\dagger)$ is the annihilation(creation) operator for the cavity mode photon and $\hat{\sigma} (\hat{\sigma}^\dagger)$ is the lowering(raising) operator for the atomic state.
The state of injected atoms is described by three parameters, $\rho_{\rm{ee}}$, $|\rho_{\rm{eg}}|$ and arg$(\rho_{\rm{eg}})$. For simplicity, we assume an uniform atom-field coupling constant $g$ and a fixed interaction time $\tau$.

Adapting the approach in Chapter 13 of Ref.\ (\textit{34})
, we assume that each atom traversing the cavity contributes to the cavity field state independently so that the gross change of the cavity density matrix $\rho_{\rm{field}}$ is just the product of an atomic injection rate $r$ 
and a single-atom contribution $\delta\rho_{\rm{field}}$, \textit{i.e.}
\begin{equation}
 \dot{\rho}_{\rm{field,interaction}}(t)= r \times \delta\rho_{\rm{field}}(t).
\end{equation}

\noindent
Each atom interacts with the cavity during the given interaction time $\tau$ and the single-atom contribution can be derived by tracing the density matrix after the interaction over the atomic state
\begin{eqnarray}
 \delta\rho_{\rm{field}}(t_0) &=& \rho_{\rm{field}}(t_0+\tau) - \rho_{\rm{field}}(t_0)\nonumber\\ 
 								&=& {\rm tr}_{\rm{atom}}[\rho_{\rm{system}}(t_0+\tau)] - \rho_{\rm{field}}(t_0).
\end{eqnarray}
\noindent
The stochastic dissipation by the cavity decay can be described by the Lindblad equation,
\begin{equation}
 \dot{\rho}_{\rm{field,decay}} = \gamma_c [2\hat{a}\rho_{\rm{field}}\hat{a}^\dagger - (\rho_{\rm{field}}\hat{a}^\dagger\hat{a} + \hat{a}^\dagger\hat{a}\rho_{\rm{field}})]
\end{equation}
\noindent
Total change in the density matrix would be the sum of these two effects. With identification of $\left<N_c\right> =r\gamma_c^{-1}$, the resulting master equation is then
\begin{eqnarray}
\dot{Q}_{n,m}/\gamma_c&=&\left<N_c\right> \{ Q_{n,m} ( \rho_{\rm{ee}} C_n C_m + \rho_{\rm{gg}} C_{n-1} C_{m-1} -1 ) \nonumber\\
 &+& Q_{n+1,m+1}\rho_{\rm{gg}}S_n S_m + Q_{n-1, m-1} \rho_{\rm{ee}} S_{n-1} S_{m-1} \nonumber\\
 &+& [i Q_{n, m+1} \rho_{\rm{eg}} C_n S_m - Q_{n+1, m} \rho_{\rm{ge}} S_n C_m \nonumber\\
 &+& Q_{n, m-1} \rho_{\rm{ge}} C_{n-1} S_{m-1} - Q_{n-1, m} \rho_{\rm{eg}} S_{n-1} C_{m-1}]\} \nonumber\\
 &+& \left[2\sqrt{(n+1)(m+1)}Q_{n+1, m+1} - (n+m) Q_{n,m}\right],\label{ME}
\end{eqnarray}
where $Q_{n,m} = \left\langle n \right|\rho_{\rm{field}}\left|m\right\rangle$, $C_n = \cos(\sqrt{n+1}g\tau)$ and $S_n = \sin(\sqrt{n+1}g\tau)$.  The steady state solution of the cavity field can be found by letting $\dot{Q}_{n,m} = 0$ in Eq.\ (11). 

\hbox{}

\noindent{\bf 2.2.1 The contribution of non-collective emission} 

When $\rho_{\rm{eg}}$ = 0, \textit{i.e.} atoms have no phase preference, the cavity field is affected only by the ground and excited state populations, $\rho_{\rm{gg}}$ and $\rho_{\rm{ee}}$. Equation (11) 
 with the steady state condition $\dot{Q}_{n,m} = 0$ is reduced to
\begin{eqnarray}
&&\left<N_c\right> [ Q_{n,m} ( \rho_{\rm{ee}} C_n C_m + \rho_{\rm{gg}} C_{n-1} C_{m-1} -1 ) \nonumber\\
&&+ Q_{n+1,m+1}\rho_{\rm{gg}}S_n S_m + Q_{n-1, m-1} \rho_{\rm{ee}} S_{n-1} S_{m-1}] \nonumber\\
&&+ \left[2\sqrt{(n+1)(m+1)}Q_{n+1, m+1} - (n+m) Q_{n,m}\right] = 0.
\end{eqnarray}
\noindent
It can be easily shown that if an initial field does not have non-zero off-diagonal density matrix elements (e.g. vacuum state) the resulting field state also does not have non-zero off-diagonal density matrix elements. So, $Q_{nn} = P_n$, the probability of having $n$ photons in the cavity, are the only remaining non-zero density matrix elements. Using the relation $C^2_n = 1 - S^2_n$ and $\rho_{\rm{gg}} = 1- \rho_{\rm{ee}}$, the recursion relation for $P_n$ can be obtained as

\begin{eqnarray}
\Big\{\left<N_c\right> \left[ P_{n+1}(1-\rho_{\rm{ee}})S^2_n - P_n \rho_{\rm{ee}} S^2_n \right]+2(n+1)P_{n+1} \Big\} \;\;\;\;\;\;\;\;\;\;\;\;\nonumber\\
- \Big\{ \left<N_c\right> \left[ P_{n}(1-\rho_{\rm{ee}})S^2_{n-1} - P_{n-1} \rho_{\rm{ee}} S^2_{n-1} \right]+2n P_n \Big\}=0.
\end{eqnarray}
%
Equation (7) is satisfied if the quantities in both curly brackets are zero for any $n$, or
\begin{equation}
\forall n:P_{n+1} =  P_n \frac{\left<N_c\right> \rho_{\rm{ee}} S^2_n}{ 2(n+1) + \left<N_c\right>( 1-\rho_{\rm{ee}})S^2_n} \;.
\end{equation}
\noindent
So, the general expression for $P_n$ is

\begin{equation}
P_n = P_0 \prod_{k=1}^n \frac{\left<N_c\right> \rho_{\rm{ee}} \sin^2(\sqrt{k}g\tau)}{ 2k + \left<N_c\right>( 1-\rho_{\rm{ee}}) \sin^2(\sqrt{k}g\tau)}
\label{P_n}
\end{equation}
and the probability $P_0$ of occupying the vacuum state can be found from the normalization condition $\sum_k P_k = 1$.

When the intracavity mean photon number $\left<n\right>$ is small enough to satisfy $\sqrt{\left<n\right>}g\tau\ll1$, the Rabi oscillation angle of each atom is not large and thus the gain saturation (see Sec.\ 2.2.4) 
does not occur. In this regime, we can approximate $\sin^2(\sqrt{k}g\tau)\simeq  k (g\tau)^2$. Then Eq.\ (15) 
 is now simplified as

\begin{equation}
P_n = P_0 \left[ \frac{p \rho_{\rm{ee}}}{2 + p (1-\rho_{\rm{ee}})}\right]^n,
\end{equation}
where the pumping parameter $p$ is defined as $p \equiv \left<N_c\right>(g\tau)^2$ and $P_0 = 1 - \rho_{\rm{ee}}p/\left[2+(1-\rho_{\rm{ee}})p\right]$.
The intracavity mean photon number $\left<n\right>$ in the steady state can then be derived as

\begin{equation}
\left<n\right> = \sum_n n P_n = \frac{\frac{1}{2}\rho_{\rm{ee}}p}{1+(1-2\rho_{\rm{ee}})\frac{1}{2}p}.
\end{equation}

It is noteworthy that the solution can be categorized in three cases depending on $\rho_{\rm{ee}}$ values. 
First, when $\rho_{\rm{ee}}>0.5$, \textit{i.e.} population inverted, the mean photon number diverges when $ p = (\rho_{\rm{ee}}-1/2)^{-1}$ under the approximation of $\sqrt{\left<n\right>}g\tau\ll1$, due to the stimulated emission amplification of the cavity field. In other words, the cavity field undergoes lasing. However, when $\rho_{\rm{ee}} < 0.5$, the absorption is larger than the stimulated emission, and therefore the photon number plateaus at $\left<n\right> = \rho_{\rm{ee}}/(1-2\rho_{\rm{ee}})$ as $p\rightarrow \infty$. Lastly, when $\rho_{\rm{ee}} = 0.5$, which means the excited and ground state populations are equal, the absorption and stimulated emission are balanced. As a result, only the spontaneous emission effect remains and the photon number is given by $\left<n\right> = 1/4\: p = 1/4\: \left<N_c\right>(g\tau)^2$, growing linearly until the gain saturation occurs.

\hbox{}

\noindent{\bf 2.2.2 The contribution of collective emission}

When the cavity is pumped by atoms in a superposition state ($\rho_{\rm{eg}} \neq 0$), the equation to solve becomes a bit complicated. Under the assumption of $\sqrt{\left<n\right>}g\tau\ll1$, we can replace $S_k$ and $C_k$ with $\sqrt{k+1}g\tau$ and 1, respectively, in the approximation up to the first order of $g\tau$. After such substitution to Eq.\ (11) 
 and keeping the terms up to the first order of $g\tau$, the equation for the steady state solution is
\begin{eqnarray}
&i \left<N_c\right>  (Q_{n, m+1} \rho_{\rm{eg}} \sqrt{m+1}g\tau - Q_{n+1, m} \rho_{\rm{ge}} \sqrt{n+1}g\tau \nonumber\\
&+ Q_{n, m-1} \rho_{\rm{ge}}  \sqrt{m}g\tau - Q_{n-1, m} \rho_{\rm{eg}} \sqrt{n}g\tau ) \nonumber\\
&+ \left[2\sqrt{(n+1)(m+1)}Q_{n+1, m+1} - (n+m) Q_{n,m}\right] = 0,
\end{eqnarray}
\noindent
which can be rearranged as
\begin{eqnarray}
 &\sqrt{n+1} ( \sqrt{m+1} Q_{n+1,m+1} - i \rho_{\rm{ge}} \left<N_c\right> g\tau  Q_{n+1,m}) \;\;\;\;\; \nonumber\\
&+ \sqrt{m+1} ( \sqrt{n+1} Q_{n+1,m+1} + i \rho_{\rm{eg}} \left<N_c\right> g\tau  Q_{n,m+1}) \;\; \nonumber\\
&- \sqrt{m} ( \sqrt{m} Q_{n,m} - i \rho_{\rm{ge}} \left<N_c\right> g\tau  Q_{n,m-1}) \;\;\;\;\;\;\;\; \nonumber\\
&+ \sqrt{n} ( \sqrt{n} Q_{n,m} + i \rho_{\rm{eg}} \left<N_c\right> g\tau  Q_{n-1,m})  =0,
\end{eqnarray}
which is satisfied if the quantity in each round bracket is zero for any $n$.
So, the general expression for $Q_{n,m}$ is then
\begin{eqnarray}
Q_{n,m} &=& i \rho_{\rm{ge}} \left<N_c\right> g\tau Q_{n,m-1}/\sqrt{m}\nonumber\\
&=& -i \rho_{\rm{eg}} \left<N_c\right> g\tau Q_{n-1,m}/\sqrt{n}\nonumber\\ 
&=& \frac{ (-i \rho_{\rm{eg}} \left<N_c\right> g\tau)^n (i \rho_{\rm{ge}} \left<N_c\right> g\tau)^m}{\sqrt{n!m!}} Q_{0,0},
\end{eqnarray}
which is exactly the density matrix element of the coherent state $\left| \alpha = -i \left<N_c\right> \rho_{\rm{eg}}  g\tau \right\rangle$.
The mean photon number of the resulting coherent state is $\left<n\right> = |\alpha|^2=(\left<N_c\right> |\rho_{\rm{eg}}| g\tau)^2$, exhibiting explicit quadratic dependence on $\left<N_c\right>$, characteristics of collective emission or the superradiance. 

Similar results were presented for different configurations by Refs.\ (\textit{17}) and (\textit{35}).
Especially, Le Kien \textit{et al.} considered a configuration with a lossless cavity. They found that the consecutive injection of $N$ atoms in the same superposition state generates a coherent state of the cavity field, $\left|\alpha = -i N \rho_{\rm eg} g\tau\right\rangle$. Comparing it with the result we derived above, we then confirm that the effective number of atoms involved in the collective emission in the single-atom superradiance is indeed $\left<N_c\right>=r/\gamma_c$.

\hbox{}

\noindent{\bf 2.2.3 Competition of two contributions}

\label{sec2-3}

In general, both effects contribute to the cavity field and the steady state mean photon number can be approximately expressed as a sum of two terms:
\begin{eqnarray}
\left<n\right> &\simeq & \frac{\left<N_c\right> \rho_{\rm{ee}}  (g\tau)^2}{2-(2\rho_{\rm{ee}}-1)\left<N_c\right> (g\tau)^2} + \left(\left<N_c\right>|\rho_{\rm{eg}}| g\tau\right)^2\nonumber\\
&\approx&  \frac{1}{2}\left<N_c\right> \rho_{\rm{ee}} (g\tau)^2 +\left<N_c\right>^2 |\rho_{\rm{eg}}|^2  (g\tau)^2, \;\;\; {\rm when} \left<N_c\right>\ll (g\tau)^{-2}.
\label{sum}
\end{eqnarray}
\noindent We have verified the solution of Eq.\ (21) 
 to be valid by comparing it with the numerical calculation results using the full density matrix equation (see Fig. S5).

The competition between two terms is independent of $g\tau$, which corresponds to the single-atom vacuum-Rabi oscillation angle, but only depends on the initial atomic state ($\rho_{\rm{ee}}, |\rho_{\rm{eg}}|$) and the atom number $\left<N_c\right>$. The second term becomes dominant when
\begin{eqnarray}
\left<N_c\right> &>& 2\rho_{\rm{ee}} / (2|\rho_{\rm{eg}}|)^2 \nonumber\\
&>& 1  \rm{\quad (for\; maximum\; dipole\; case,\; |\rho_{\rm{eg}}| = 1/2)}
\label{lbound}
\end{eqnarray}

\hbox{}

\noindent{\bf 2.2.4 Saturation due to coherent Rabi oscillation}

\label{sec2-4}
As in the cavity-QED microlaser, when the accumulated cavity photon number is sufficiently large, the Rabi angle $\sqrt{\left<n\right>}g\tau$ becomes large and thus the coherent Rabi oscillation of each atom gives a feedback to the photon number. As a result, the photon number becomes stabilized (\textit{16,17}). 
This process starts when the Rabi angle due to the cavity field becomes comparable to the inversion angle of atoms (given by the pump pulse area $\Theta$ in the experiment).
\begin{equation}
\sqrt{\left<n\right>+1}g\tau \sim  \Theta
\end{equation}
Assuming that the dipole pumping is dominant [Eq.\ (22)]
, we substitute $\left<n\right>$ with the result from Sec.\ 2.2.3. 
 The saturation condition is then
\begin{equation}
	\left<N_c\right> \sim (g\tau)^{-2} \frac{\Theta}{\sin\Theta} \sim (g\tau)^{-2}.
\end{equation}

\hbox{}

\noindent{\large \bf 2.3 Connection to thresholdless lasing}

\hbox{}

Conventionally, the thresholdless lasers are studied in terms of the so-called $\beta$ factor, which is the ratio of the spontaneous emission rate into the lasing cavity mode to the total decay rate (radiative as well as non-radiative) of the gain medium (of atoms or molecules). 
Specially designed micro/nano-cavities are used to maximize $\beta$ factor by means of the Purcell effect in order to lower the lasing threshold.
Thresholdless lasing is achieved when $\beta \simeq 1$, for which almost all energy supplied to the excited state of the gain medium is transferred to the cavity mode per unit time via spontaneous emission of photons ({\it 24}).


Let us consider $\beta$ factor in the non-collective emission case (usual lasing) in our experiment. 
From Eq.~(21), with $\rho_{\rm eg}=0$
, we get the output flux of photons as $2\gamma_c \langle n\rangle\simeq  \gamma_c\langle N_c\rangle \rho_{\rm ee} (g \tau)^2$, which should be balanced by the emission from the atom to the cavity mode. Therefore, the spontaneous emission rate into the cavity mode is $\gamma_c\langle N_c\rangle \rho_{\rm ee} (g \tau)^2$. Total decay rate of the excited state energy must be equal to the total pumping rate of energy in the excited state, $\langle N\rangle \rho_{\rm ee}/\tau=\gamma_c\langle N_c\rangle \rho_{\rm ee} $ using the relation $\langle N\rangle=\langle N_c \rangle \gamma_c\tau$. From this consideration, we then obtain $\beta=(g\tau)^2$.

In the non-collective emission experiment of Fig.~3B, $\beta\simeq 0.011$.
Despite the large single-atom cooperativity of $C\simeq44$, most of the excited atoms exit the cavity without emitting a photon into the cavity because of the short interaction time $\tau$.
This is consistent with the large mean photon number change occurring at the threshold in Fig.~3B.
On the other hand, if we just apply this definition to the case of Fig.~3A, $\beta \simeq 0.034$, which certainly cannot explain the thresholdless growth of the mean photon number observed there.

One may extend the definition of $\beta$ factor to the collective emission case by interpreting the spontaneous emission into the cavity mode to include the collective emission also. The collective emission rate into the cavity mode is obtained from Eq.~(21) as $2\gamma_c\langle N_c\rangle^2 |\rho_{\rm eg}|^2 (g\tau)^2$, and thus for collective emission we get $\beta_{\rm coll}=2\langle N_c\rangle (|\rho_{\rm eg}|^2/\rho_{\rm ee}) (g\tau)^2$. 
As the atom number $\langle N_c \rangle$ increases, $\beta_{\rm coll}$ also increases and finally converges to unity as the saturation condition, Eq.\ (24), sets in. So, it is tempting to use this effective $\beta$ factor for explaining the thresholdless lasing in Fig.~3A.
However, this effective $\beta$ is not constant. It depends on the atom number $\langle N_c\rangle$ or pumping rate, and thus no longer a geometric factor, loosing its usefulness in analysis. This consideration led us to conclude that the thresholdless lasing coming from the collective emission or superradiance in Fig.~3A cannot be explained in terms of the usual $\beta$ factor in ordinary lasers.

\clearpage
\setcounter{figure}{0}
\makeatletter 
\renewcommand{\thefigure}{S\@arabic\c@figure}
\makeatother

\begin{figure}
	\center \includegraphics[width=1\textwidth]{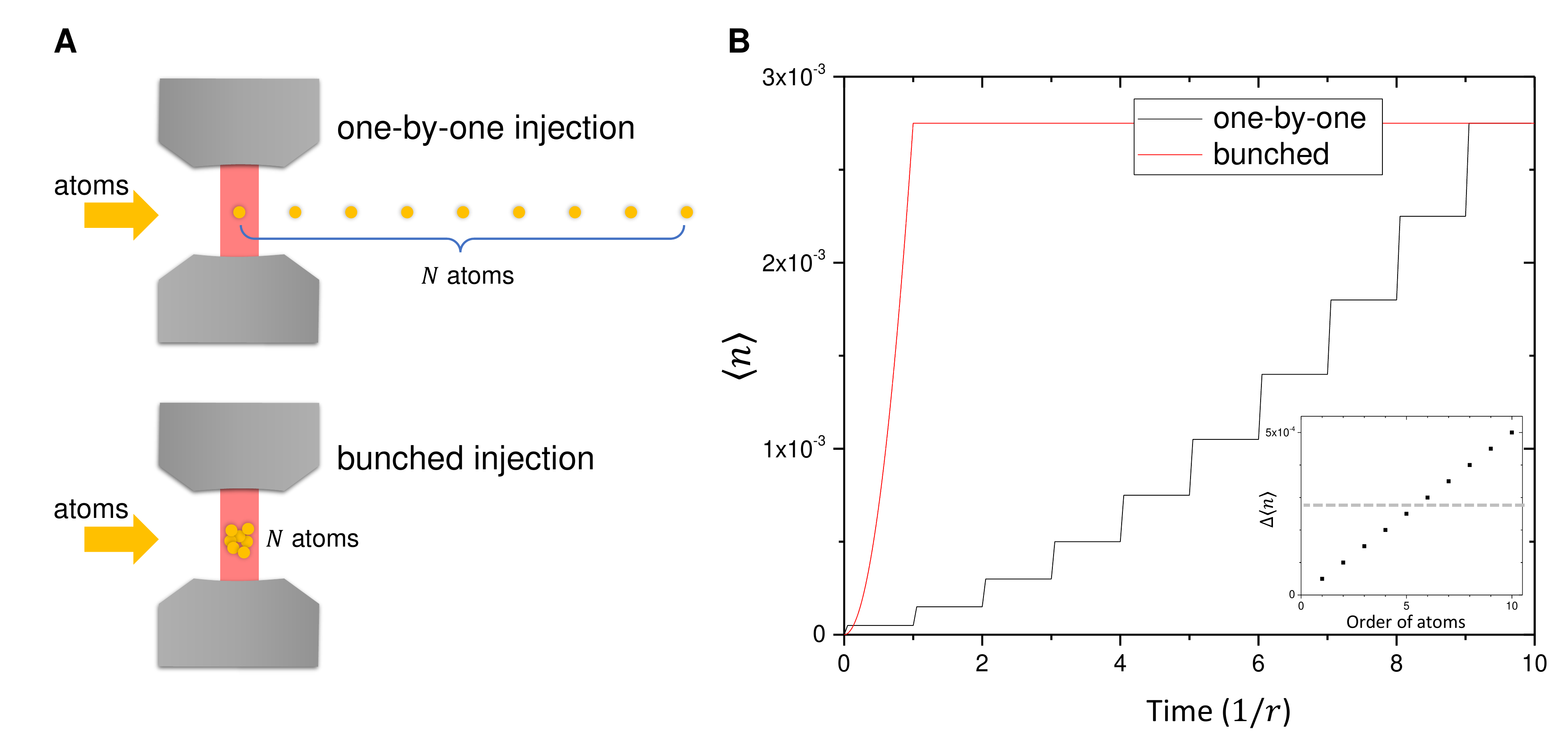}
	\caption{
		\textbf{Comparison between one-by-one and bunched atomic injections in a lossless cavity.}
		({\bf A}) For one-by-one injection we assume that the atoms (yellow circles) enter the cavity mode (red-shaded area) separately with a time interval of $1/r$, where $r$ is the atomic injection rate. 
		On the other hand, for bunched injection, we assume that the whole number of atoms enters the cavity at once and interact with the cavity for a time period of $1/r$, as in the usual superradiance.
		({\bf B}) Mean intracavity photon number is plotted for both cases as a function of time in the unit of the injection time interval $1/r$. The atoms are prepared in the superposition state $\left|\psi_{\rm atom}\right\rangle = (\left|{\rm e}\right\rangle + \left|{\rm g}\right\rangle)/\sqrt{2}$ before injection. 
		For one-by-one injection, the atom-field interaction time $\tau$ is assumed to be much shorter than $1/r$ with $g\tau = 0.01$. In this case,
		$\left<n\right>$ increases stepwise as the single atoms traverse the cavity. 
		After the same number of atoms (10 in the plot) has gone through the cavity, the resulting $\left<n\right>$ is the same for both injection cases. 
		The step increment $\Delta\left<n\right>$ indicates the photo-emission amount of each atom. It linearly increases as the number of the preceding atoms grows due to the one-sided field-mediated interaction (inset). 
		As a result, the photo-emission of the last atom, interacting with the preceding $N-1$ atoms, is approximately twice stronger than the average photo-emission (horizontal dashed line) in the bunched injection case.
	}
	\label{exfig:bunch}
\end{figure}

\begin{figure}
	\center \includegraphics[width=0.95\textwidth]{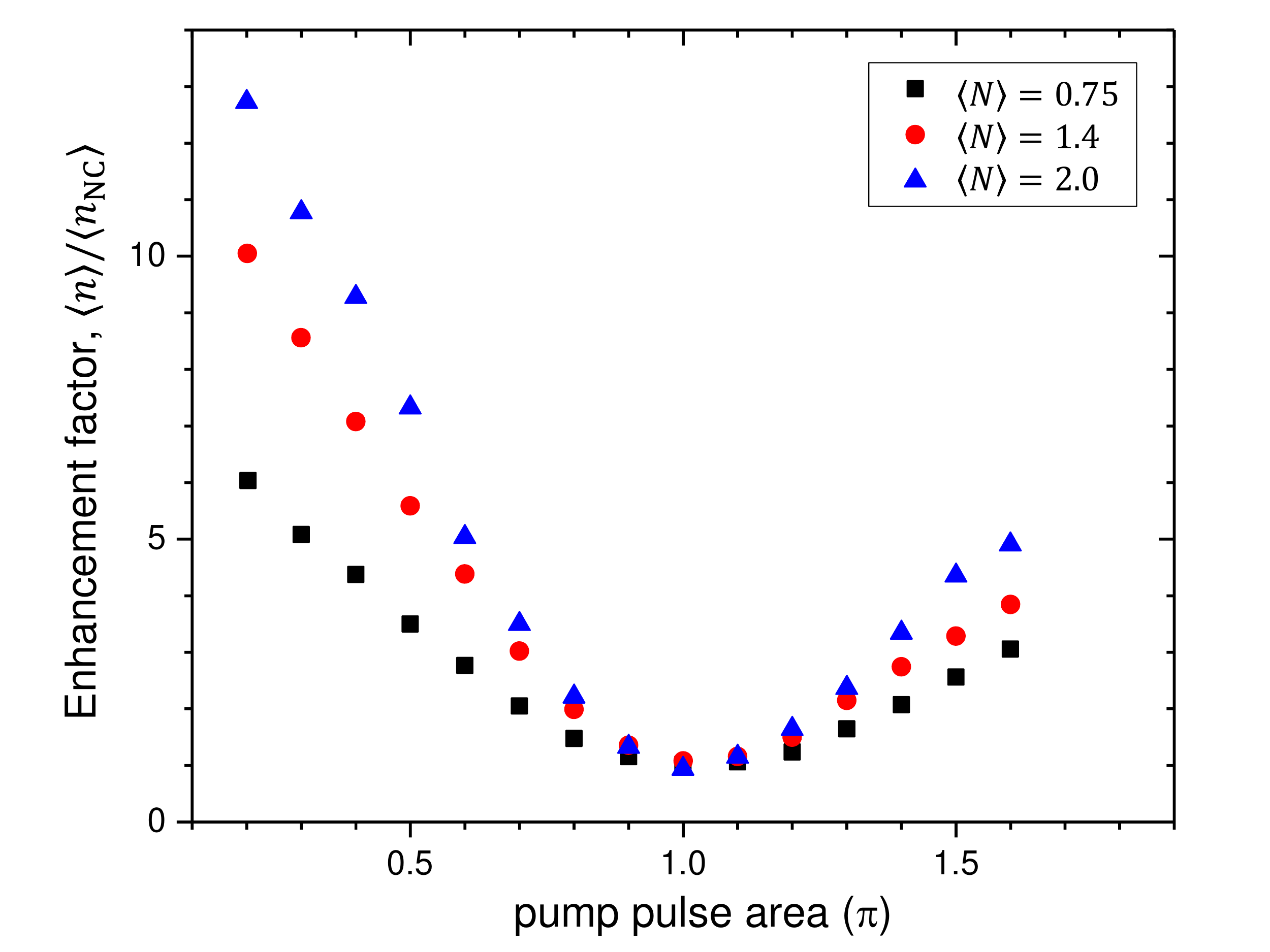}
	\caption{
		\textbf{Enhancement factor in the single-atom superradiance.}
		We define the enhancement factor as $\left<n\right>/\left<n_{\rm NC}\right>$, where $\left<n\right>$ is the mean photon number observed with phase-aligned atoms in Fig.\ 2A of the main text and $\left<n_{\rm NC}\right>$ is the expected mean photon number (dashed curves in Fig.\ 2A of the main text) by the non-collective emission alone, \textit{i.e.} $\rho_{\rm{eg}} = 0$. The enhancement factor even exceeds 10 for small pump pulse areas.
	}
	\label{exfig:enhancement}
\end{figure}

\begin{figure}
	\center \includegraphics[width=0.9\textwidth]{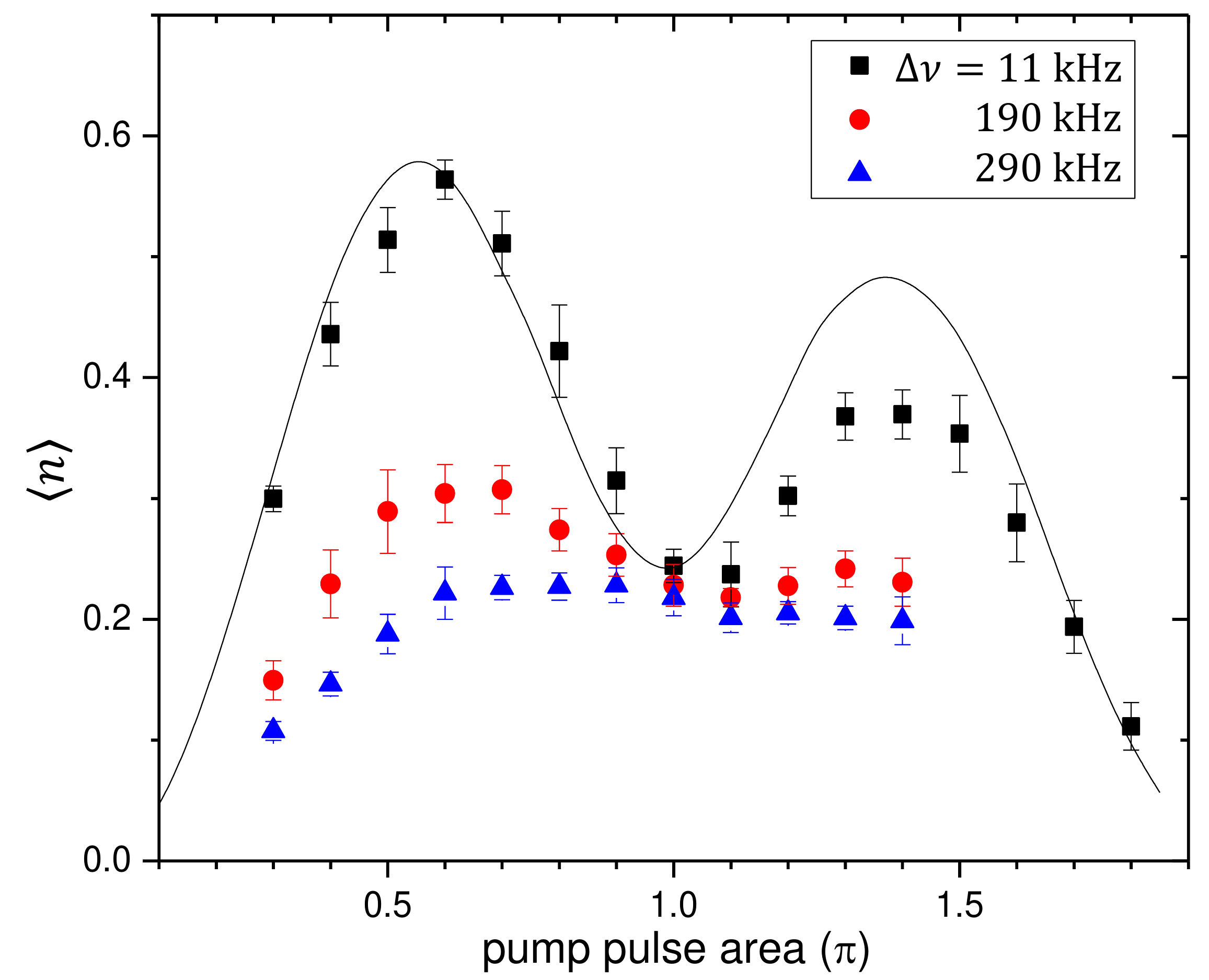}
	\caption{
		\textbf{Effect of pump laser phase diffusion on the single-atom superradiance.}
		By intentionally degrading the frequency stabilization of the pump laser, we can induce additional phase noise in the pump laser and can experimentally verify the effect of pump phase diffusion on the single-atom superradiance. 
		Here $\Delta\nu$ denotes the root-mean-square linewidth of the pump laser. The larger the laser linewidth is, the smaller enhancement is achieved by the single-atom superradiance. The measured intracavity atom number was 0.57. The solid line is a theoretical prediction based on QTS.
	}
	\label{exfig:linewidth}
\end{figure}

\begin{figure}
	\center \includegraphics[width=1\textwidth]{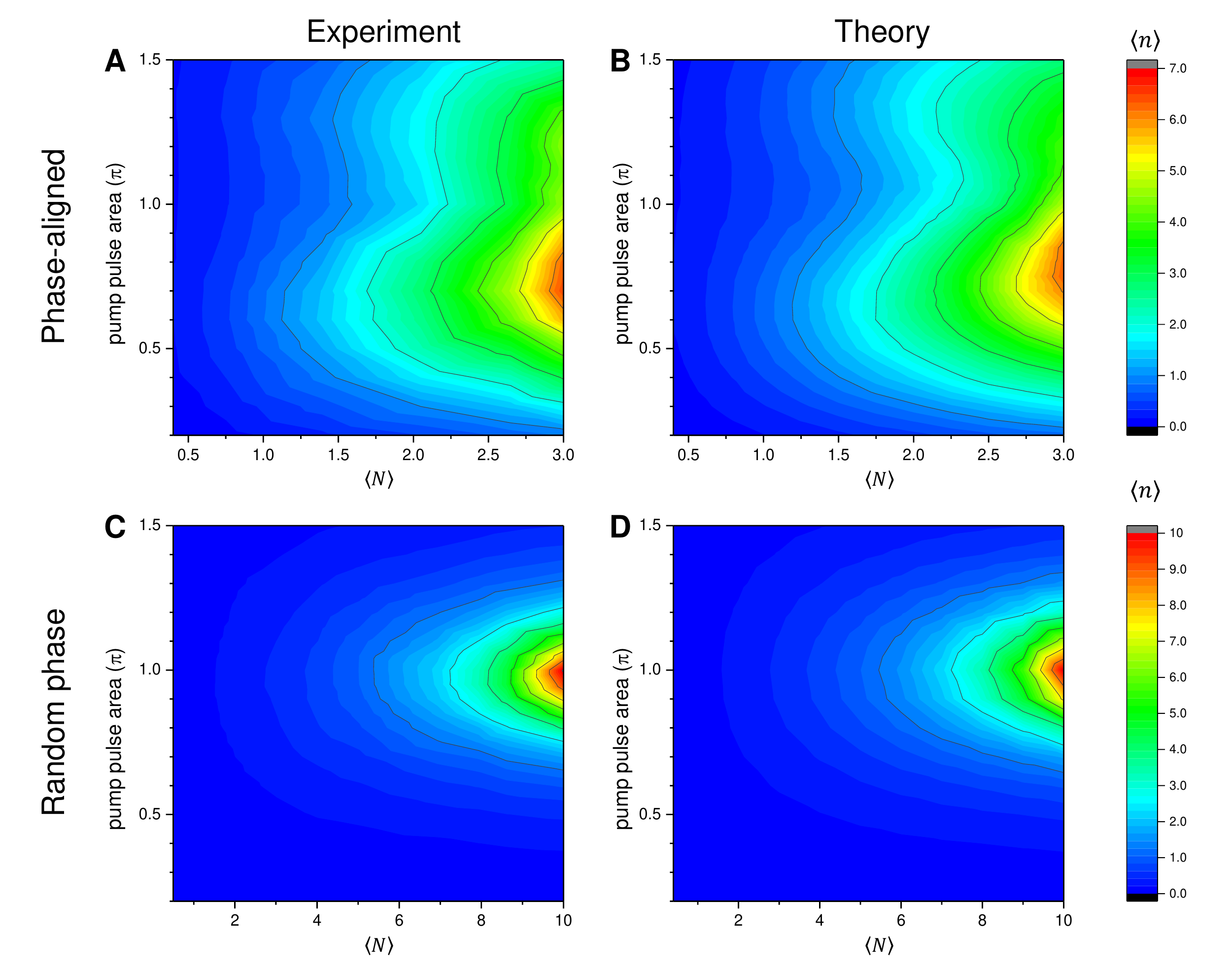}
	\caption{
		\textbf{Dependence on the pump pulse area $\Theta$ and the intracavity mean atom number $\left<N\right>$ of the intracavity mean photon number $\left<n\right>$.}
		Density plot of $\left<n\right>$ as a function of $\Theta$ and $\left<N\right>$.  Experimental results ({\bf A}, {\bf C}) are compared with the numerical simulation results ({\bf B}, {\bf D}). Plots in {\bf A} and \textbf{B} (\textbf{C} and \textbf{D}) share the same color scale.
	}
	\label{exfig:density}
\end{figure}

\begin{figure}
	\center \includegraphics[width=0.9\textwidth]{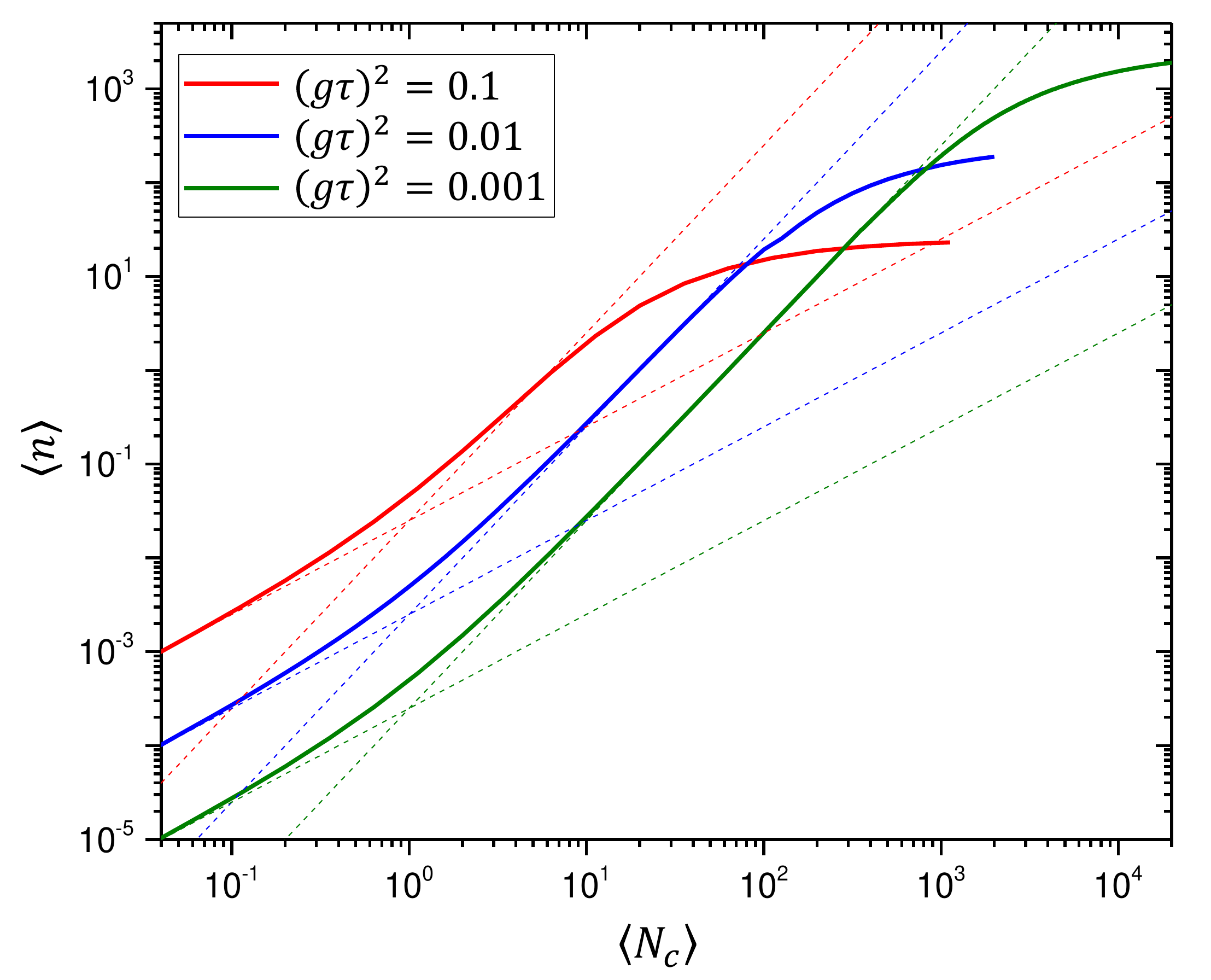}
	\caption{
		\textbf{Region of quadratic dependence.}
		Based on the quantum master equation, the intracavity mean photon number $\left<n\right>$ in the case of atoms prepared in $\left|\psi_{\rm atom}\right\rangle = (\left|{\rm e}\right\rangle + \left|{\rm g}\right\rangle)/\sqrt{2}$ is plotted as a function of $\left<N_c\right>$ for various $g\tau$ values. Dashed lines with log-log slopes of 1 and 2 indicate approximative solutions for the cases of non-collective and collective emission alone, respectively, and solid lines are solution of the full-density-matrix master equation. The numerical solutions well match the sum of two approximative solutions expect for the saturation at $\left<N_c\right>>(g\tau)^{-2}$. Note that the transition of the log-log slope from 1 to 2 occurs near $\left<N_c\right> = 1$. Furthermore, for $\left<N_c\right>>(g\tau)^{-2}$, coherent Rabi oscillation kicks in and strongly suppresses photon number fluctuation, leading to photon number saturation.  The cavity field in this region exhibits non-classical photon statistics of sub-Poissonian (\textit{16}).
	}
	\label{exfig:dipole}
\end{figure}

\begin{figure}
	\center \includegraphics[width=0.9\textwidth]{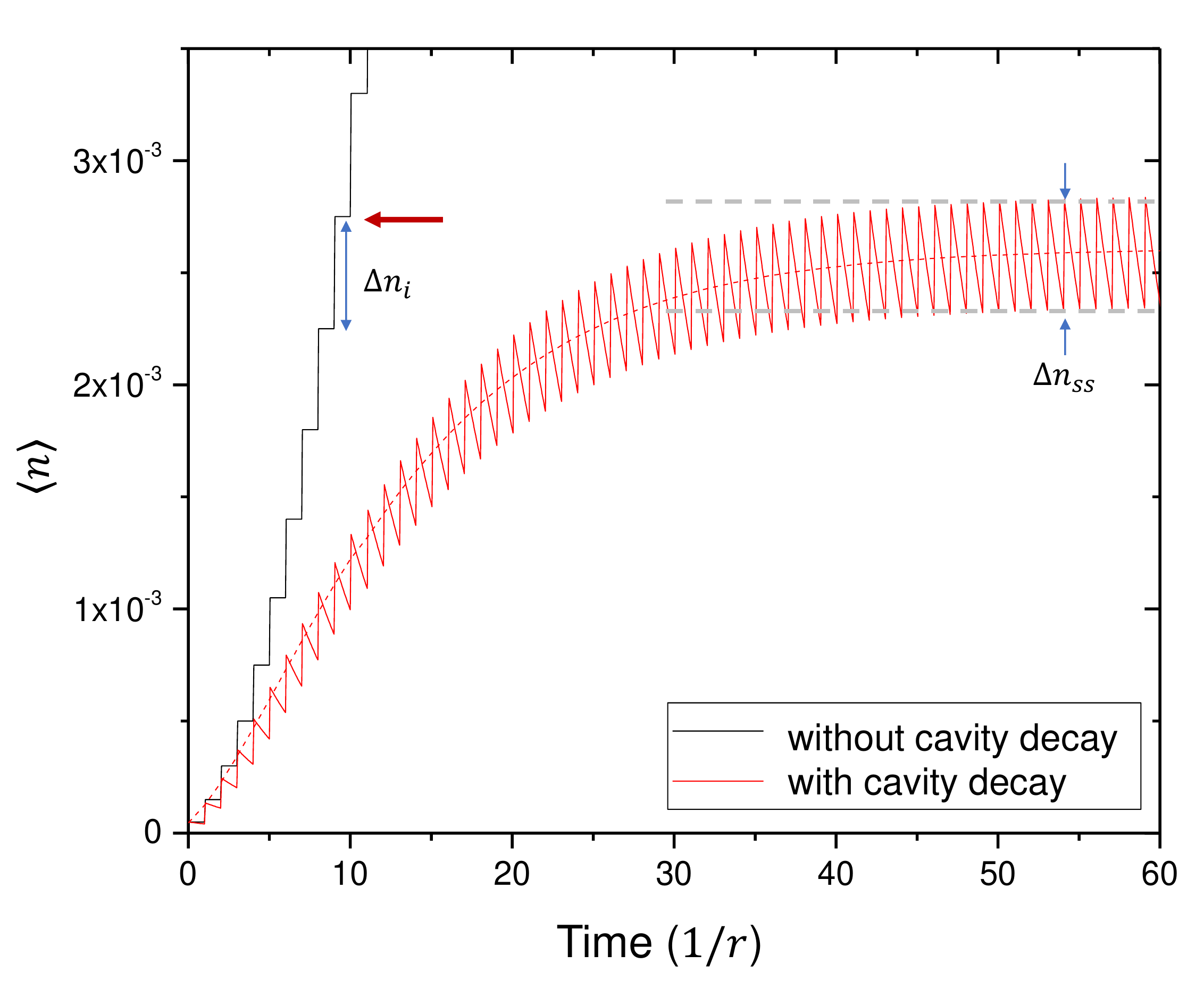}
	\caption{		
		\textbf{Transient dynamics of the single-atom superradiance.}
		With and without cavity decay, the intracavity photon number obtained from numerical simulation is plotted as a function of time. The atoms are prepared in the superposition state $\left|\psi_{\rm atom}\right\rangle = (\left|{\rm e}\right\rangle + \left|{\rm g}\right\rangle)/\sqrt{2}$ and regularly injected into the cavity mode one by one with a time interval of $1/r$, where $r$ is the atomic injection rate. 
		The atom-field interaction time $\tau$ is assumed to be much shorter than $1/r$ with $g\tau = 0.01$. 
		We adjusted the cavity decay rate in such a way that the number $\left<N_c\right>$ of atoms traversing the cavity during the cavity field decay time equals 10.
		In the presence of the cavity decay, the cavity field reaches its steady state satisfying the gain-loss balance and the converged $\left<n\right>$ is approximately equal to the mean photon number in a lossless cavity after $\left<N_c\right>$ atoms has traversed the cavity (indicated by a red arrow). Moreover, the emission $\Delta n_{ss}$ by each passing atom is also equal to the emission increment $\Delta n_i$ (with $i=\left<N_c\right>$) by the $\left<N_c\right>$-th atom in a lossless cavity,
		which is twice stronger than the emission per atom in the usual superradiance as shown in Fig. S1.
	}
	\label{exfig:chiral}
\end{figure}

\end{document}